\documentstyle[lump,psfig]{article}
\begin{document}
\begin{flushleft}
{\Large\bf Lattice Models of Quantum Gravity}\\
~\\
~\\
{\large\it E. Bittner, A. Hauke, H. Markum, J. Riedler \footnote{\small
Supported in part by {\it Fonds zur F\"orderung der wissenschaftlichen 
Forschung} under Contract P11141-PHY.}}\\[1mm]
{\small Institut f\"ur Kernphysik, Technische Universit\"at Wien,
A-1040 Vienna, Austria}\\
~\\
{\large\it C. Holm}\\[1mm]
{\small Max-Planck-Institut f\"ur Polymerforschung, D-55128 Mainz,
Germany}\\
~\\
{\large\it W. Janke \footnote{\small W. J. thanks the {\it Deutsche
Forschungsgemeinschaft} for a Heisenberg Fellowship.}}\\[1mm]
{\small Institut f\"ur Physik, Johannes Gutenberg-Universit\"at,
D-55099 Mainz, Germany}\\
~\\
~\\
\end{flushleft}
\noindent
\small
{\bf Abstract.} In Standard Regge Calculus the quadratic link lengths of the
considered simplicial manifolds vary continuously, whereas in the 
$Z_2$-Regge Model they are restricted to two possible values. The goal 
is to determine whether the computationally more easily accessible $Z_2$ 
model retains the characteristics of standard Regge theory. We study both 
models in two dimensions employing the same functional integration measure 
and determine their phase structure by Monte Carlo simulations and 
mean-field theory.\\
~\\
~\\
\normalsize
Starting point for both Standard Regge Calculus (SRC) and the $Z_2$-Regge 
Model ($Z_2$RM) is Regge's discrete description of General Relativity in 
which space-time is represented by a piecewise flat, simplicial manifold:
the Regge skeleton.\cite{regge,nonpert} The beauty of this procedure is 
that it works for any space-time dimension $d$ and for metrics of arbitrary 
signature. The Einstein-Hilbert action translates into
\begin{equation} \label{Iq}
I(q) = \lambda\,\sum_{s^d}\,V(s^d) - 2\beta\,\sum_{s^{d-2}}\delta(s^{d-2})
V(s^{d-2})\quad,
\end{equation}
with the quadratic edge lengths $q$ describing the dynamics of the lattice,
$\lambda$ being the cosmological constant, and $\beta$ the bare Planck mass 
squared.
The first sum runs over all $d$-simplices $s^d$ of the simplicial complex and
$V(s^d)$ is the $d$-volume of the indicated simplex. The second term describes
the curvature of the lattice, that is concentrated on the $(d-2)$-simplices
leading to deficit angles $\delta(s^{d-2})$, and is proportional to the
integral over the curvature scalar in the classical Einstein-Hilbert action
of the continuum theory.

In two dimensions this is easily illustrated by choosing a triangulation of the 
surface under consideration. Quantization of SRC proceeds by evaluating the 
Euclidean path integral
\begin{equation} \label{ZSRC}
Z=\left[\prod _l\int_0^\infty dq_lq_l^{-m}{\cal F}(\{q_l\})\right]
\exp(-\lambda\sum_t A_t+2\beta\sum_i\delta_i)~.
\end{equation}
In principle the functional integration should extend over all metrics on
all possible topologies, but, as is usually done, we restrict ourselves to
one specific topology, the torus. In two dimensions the sum over the deficit
angles per vertex $\delta_i$ representing the scalar curvature corresponds 
due to the Gauss-Bonnet theorem to a vanishing Euler characteristic
$2\pi\chi=\sum_i\delta_i=0$. Consequently the action in the exponent of
Eq.~(\ref{ZSRC}) consists only of a cosmological constant $\lambda$ times
the sum over all triangle areas $A_t$. The path-integral approach
suffers from a non-uniqueness of the integration measure and it is even
claimed that the true measure is of non-local nature.\cite{meno} We used
as a trial functional integration measure the expression within the square
brackets of Eq.~(\ref{ZSRC}) with $m\in I\!\!R$ permitting to investigate
a 1-parameter family of measures. The function ${\cal F}$ constrains
the integration to those configurations of link lengths which do not
violate the triangle inequalities.

Although SRC can be efficiently vectorized for large scale computing, it 
is still a very time demanding enterprise. One therefore seeks for suitable
approximations which will simplify the SRC and yet retain most of its 
universal features. The $Z_2$RM could be such a desired 
simplification.\cite{z2rm} Here the quadratic link lengths of the simplicial 
complexes are restricted to take on only the two values 
\begin{equation} \label{q}
q_l=1+\epsilon\sigma_l~,~~0<\epsilon <\epsilon_{max}~,~~\sigma_l =\pm 1~,
\end{equation} 
in close analogy to the ancestor of all lattice models, the Ising-Lenz model.
Thus the area of a triangle with edges $q_1, q_2, q_l$ is expressed as 
\begin{eqnarray} \label{at}
A_t &=&\frac{1}{2}\left| \begin{array}{cc}
       q_1 & \frac{1}{2}(q_1+q_2-q_l) \\
       \frac{1}{2}(q_1+q_2-q_l) & q_2
       \end{array} \right|^\frac{1}{2} = \nonumber \\[2mm]
    &=&c_0+c_1(\sigma_1+\sigma_2+\sigma_l)+c_2(\sigma_1\sigma_2 +
 \sigma_1\sigma_l+\sigma_2\sigma_l)+c_3\sigma_1\sigma_2\sigma_l ~.
\end{eqnarray}
The coefficients $c_i$ depend on $\epsilon$ only and impose the condition 
$\epsilon <\frac{3}{5}=\epsilon_{max}$ in order to have real and positive 
triangle areas,\cite{z2rm} i.e. \hbox{${\cal F}=1$} for all possible 
configurations. The measure $\prod_l\int dq_lq_l^{-m}$ in Eq.~(\ref{ZSRC})
is replaced by
\begin{equation}
\sum\limits_{\sigma_l = \pm 1}\exp[-m\sum_l\ln(1+\epsilon\sigma_l)]=
\sum\limits_{\sigma_l = \pm 1}\exp[-N_1m_0(\epsilon) -
\sum_l m_1(\epsilon)\sigma_l]~,
\end{equation}
where $N_1$ is the total number of links, 
$m_0 = -\frac{1}{2}m\epsilon^2 + O(\epsilon^4)$, 
$m_1 = m[\epsilon + \frac{1}{3}\epsilon^3 + O(\epsilon^5)]=mM$ and
\hbox{$M=\sum_{i=1}^{\infty}\frac{\epsilon^{2i-1}}{2i-1}$}.
Hence the path integral Eq.~(\ref{ZSRC}) translates for the $Z_2$RM into
\begin{equation} \label{Z2RM}
Z=\!\sum_{\sigma_l=\pm 1}J\exp\{-\sum_l(2\lambda c_1 + m_1)\sigma_l-
\lambda\sum_t[c_2(\sigma_1\sigma_2 + \sigma_1\sigma_l+\sigma_2\sigma_l)+
c_3\sigma_1\sigma_2\sigma_l]\},
\end{equation}
with an unimportant constant $J$. If we view $\sigma_l$ as a spin variable
and assign it to the corresponding link $l$ of the triangulation then 
$Z$ reads as the partition function of a spin system with two- and three-spin 
nearest neighbour interactions on a Kagom{\'e} lattice.
A particular simple form of Eq.~(\ref{Z2RM}) is obtained if 
$m_1=-2\lambda c_1$ and therefore 
\begin{equation}
m=\frac{-2\lambda c_1}{M} ~,
\end{equation}
which is henceforth used for the measure in the $Z_2$RM as well as in SRC.
We set the parameter $\epsilon=0.5$ in the following.

\begin{figure}
\centerline{\hbox{\small Standard Regge Calculus \hspace{30mm}
 $Z_2$-Regge Model}}
\vspace{-1mm}
\centerline{\hbox{
\psfig{figure=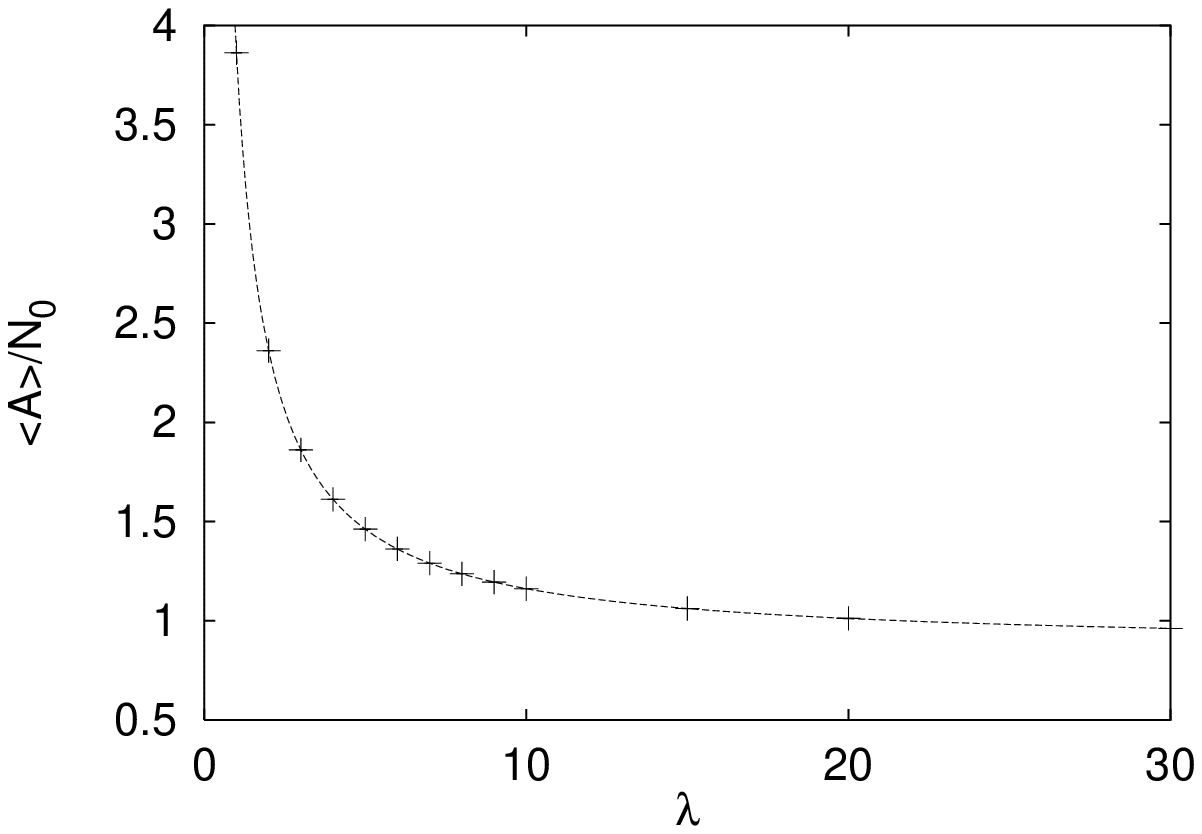,height=4.25cm,width=6cm}
\psfig{figure=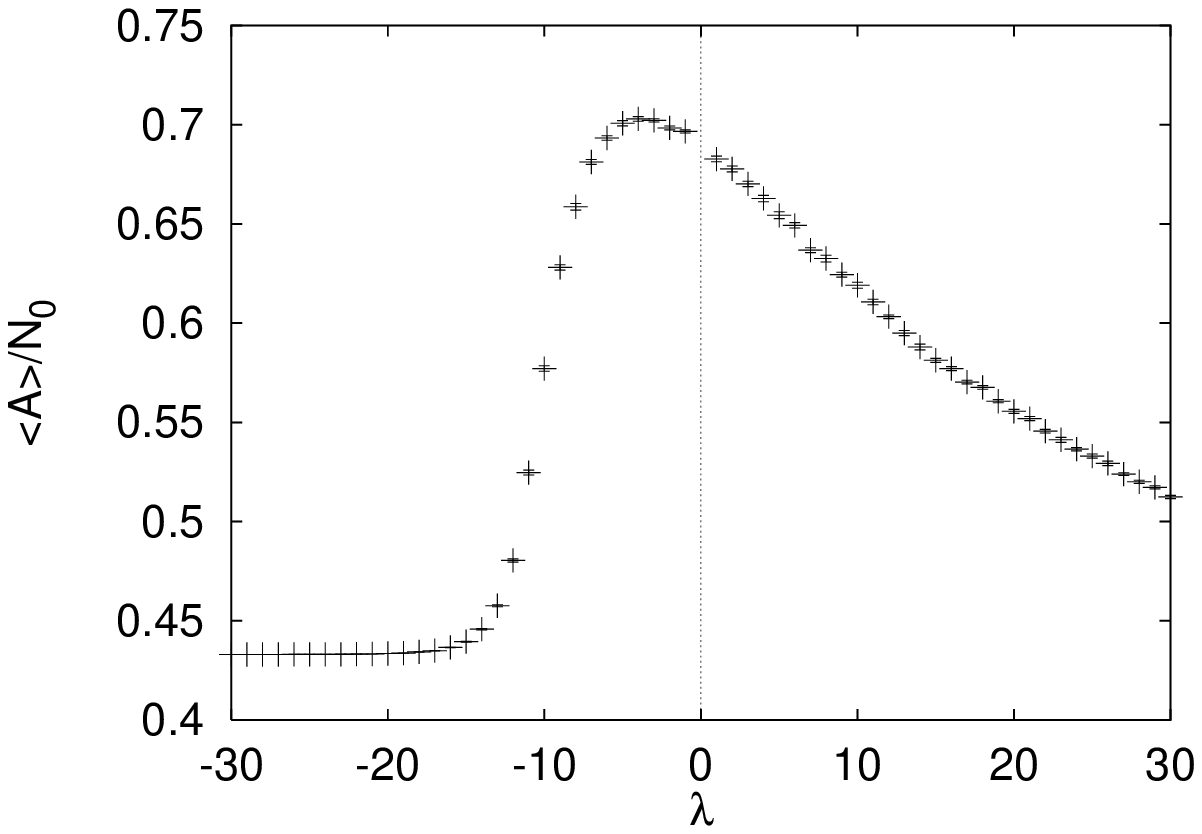,height=4.25cm,width=6cm}
}}
\centerline{\hbox{
\psfig{figure=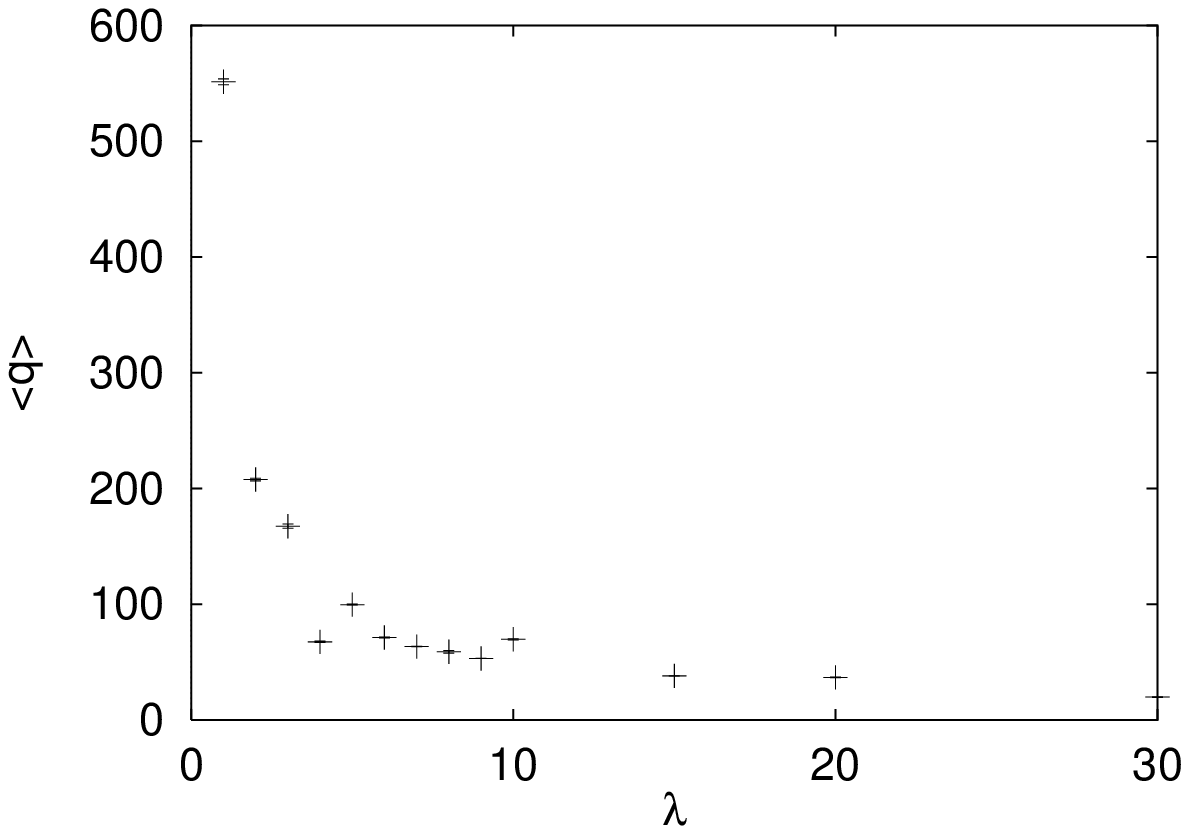,height=4.25cm,width=6cm}
\psfig{figure=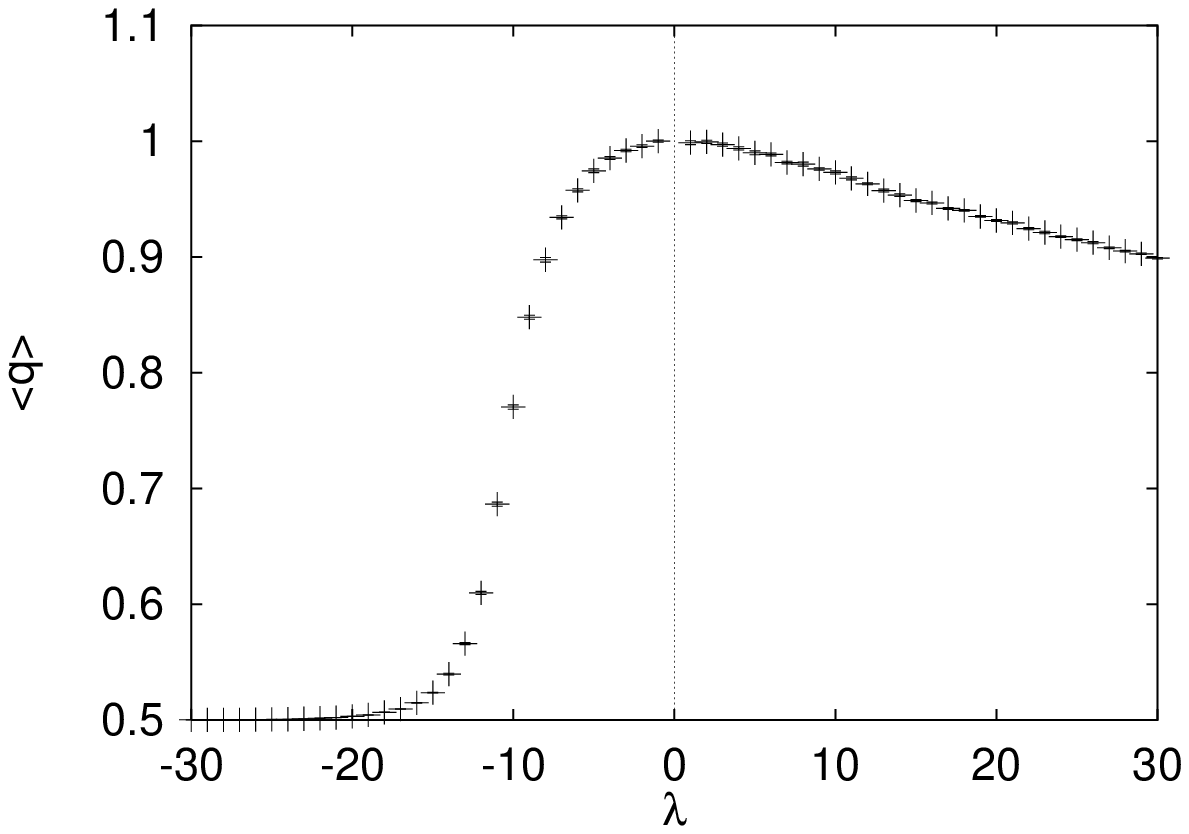,height=4.25cm,width=6cm}
}}
\centerline{\hbox{
\psfig{figure=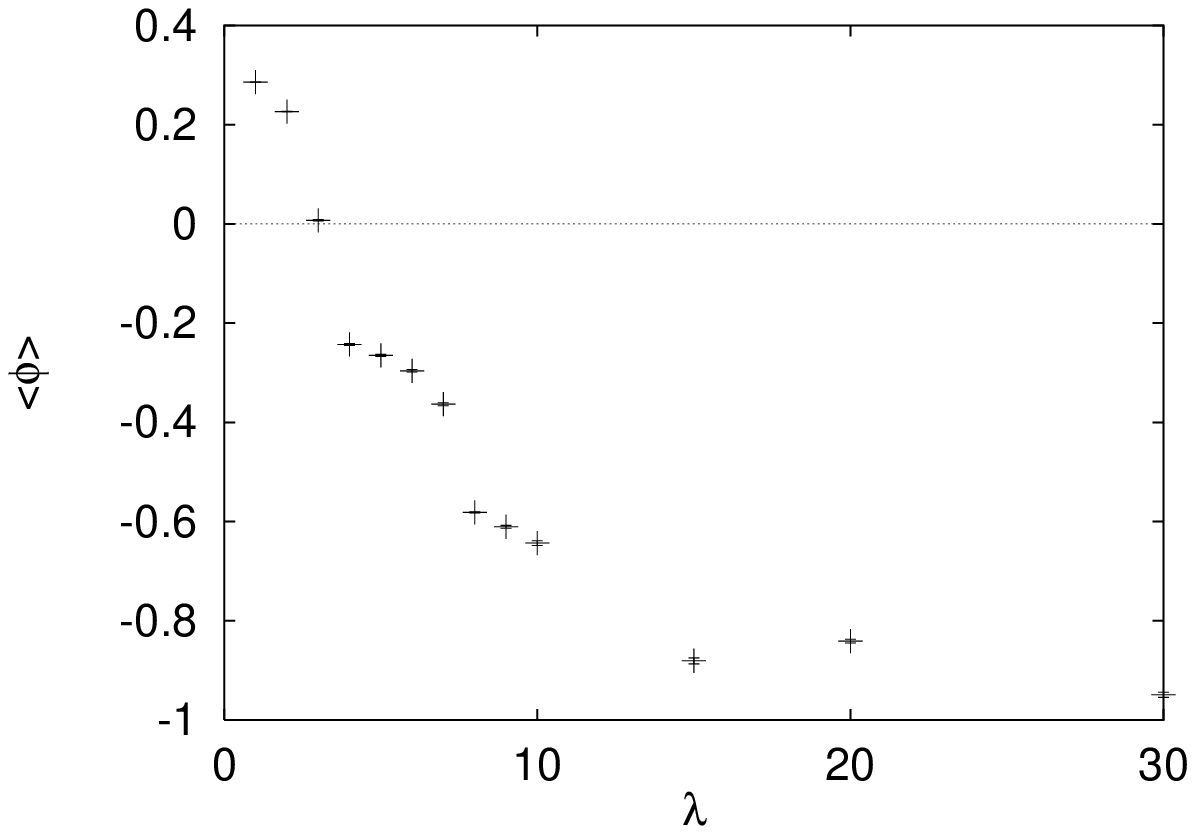,height=4.25cm,width=6cm}
\psfig{figure=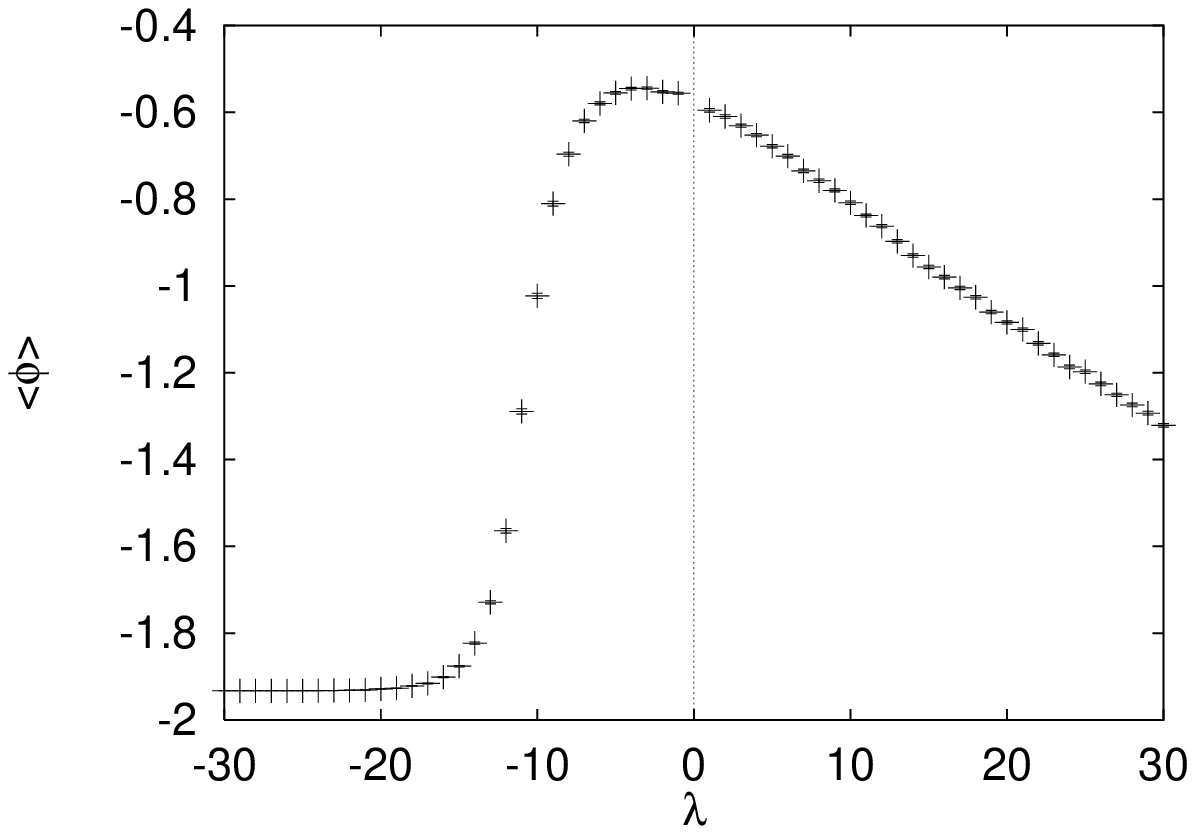,height=4.25cm,width=6cm}
}}
\centerline{\hbox{
\psfig{figure=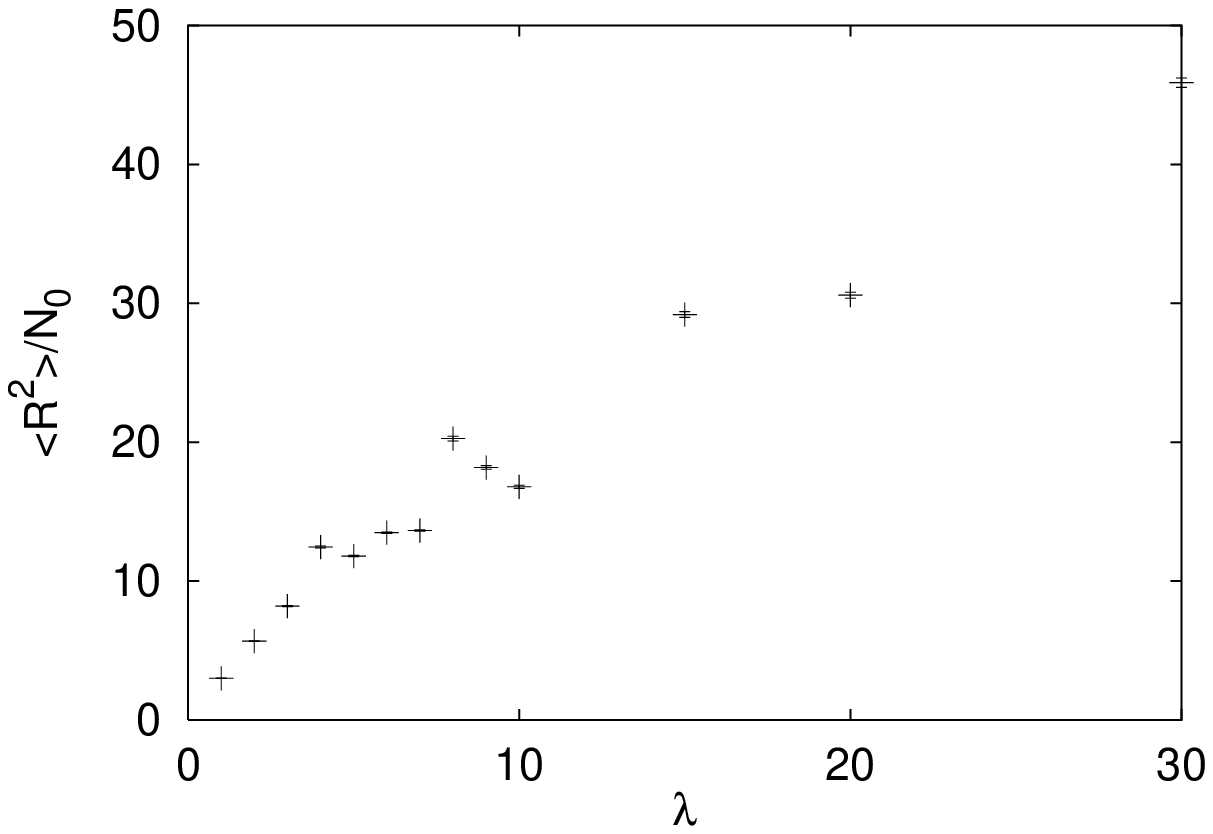,height=4.25cm,width=6cm}
\psfig{figure=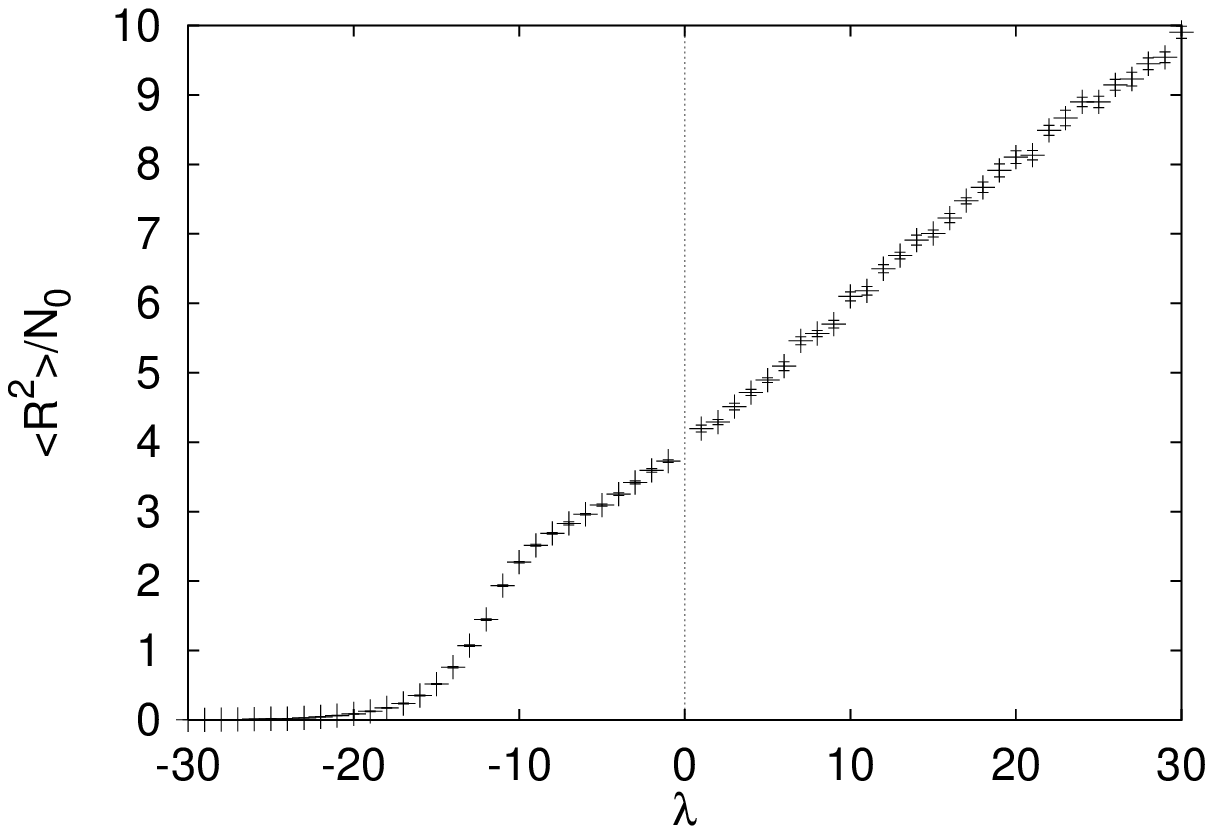,height=4.25cm,width=6cm}
}}
\vspace{3mm}
{\small {\bf Fig.~1.} Expectation values of the area $A$, the average
squared link length $q$, the Liouville field $\phi$, and the squared
curvature $R^2$ as a function of the cosmological constant $\lambda$
for the Standard Regge Calculus (left plots) and the $Z_2$-Regge Model
(right plots). $N_0$ is the total number of vertices.}
\end{figure}

To compare both models we examined the quadratic link lengths and the area 
fluctuations on the simplicial lattice. Furthermore the Liouville mode is of
special interest because it represents the only degree of freedom of pure 
2d-gravity. The discrete analogue of the continuum Liouville field 
$\varphi(x)=\ln\sqrt{g(x)}$ is defined by 
\begin{equation}
\phi=\frac{1}{A}\sum_i\ln A_i~,~~~A_i=\frac{1}{3}\sum_{t\supset i}A_t~,
\end{equation}
where $A_i$ is the area element of site $i$ and $A$ the total area.\cite{ham}
Additionally we are interested in the squared curvature defined by 
\vspace{-1mm}
\begin{equation}
R^2=\sum_i\frac{\delta_i^2}{A_i}~.
\end{equation} 
Figure~1 displays the corresponding expectation values as a function of the 
cosmological constant $\lambda$ measured from 100k Monte Carlo sweeps after 
thermalization on simplicial lattices with $16\times 16$ vertices.

Within the SRC the area increases with decreasing $\lambda$ in perfect 
agreement with the scaling relation 
\begin{equation}
\langle A\rangle = N_1\frac{1-m}{\lambda} ~.
\end{equation}
One also expects that $\langle q\rangle$ will increase as $\lambda$ tends 
to zero. Actually we observe that the system thermalizes extremely slowly 
for very small $\lambda$ and therefore display only statistically reliable 
data points for $\lambda \ge 1$ in the plots on the l.h.s. of Fig.~1. The 
Liouville field $\langle \phi \rangle$ behaves accordingly, and the 
expectation value of $R^2$ increases with $\lambda$.

Whereas the SRC becomes ill-defined for negative couplings $\lambda$, the
$Z_2$RM as an effective spin system is well-defined for all values of the
cosmological constant. The phase transition the $Z_2$RM undergoes at 
$\lambda_c\approx -11$ can be viewed as the relic of the transition from 
a well- to an ill-defined regime of SRC. A mean-field calculation for the 
$Z_2$RM to extract the critical cosmological constant leads to 
$\lambda_c=-7.012$ and gives evidence for a (weak) first-order phase 
transition. It is a well-known property of mean-field theory to 
underestimate the critical coupling. The ratio between the mean-field and 
the numerical value of $\lambda_c$ is in the same order of magnitude as 
for the exactly solvable, two-dimensional Ising model. A first-order
phase transition would prevent to gain important information about the
continuum theory, however, the true nature of the phase transition still
remains to be determined.

Another interesting quantity to consider would be the Liouville susceptibility
\begin{equation}
\chi_\phi = \langle A \rangle [\langle \phi^2 \rangle 
                             - \langle \phi \rangle^2] ~.
\end {equation}
From continuum field theory it is known that for fixed total area $A$ the
susceptibility scales according to
\begin{equation}
\ln\chi_{\phi}(L) \stackrel{L\to\infty}{\sim} c +
(2 - \eta_{\phi}) \ln L ~,
\end{equation}
with $L=\sqrt{A}$ and the Liouville field critical exponent $\eta_{\phi}=0$.
This has indeed been observed for SRC with the $dq/q$ scale invariant measure
and fixed area constraint.\cite{ham} It is, however, a priori not clear if
this feature will persist in the present model due to the fluctuating area and
the non-scale invariant measure. This point is presently under investigation.

To conclude, physical observables like the Liouville field and the squared
curvature behave similar in the $Z_2$RM and SRC for the bare coupling 
$\lambda>\lambda_c$. The phase transition of the $Z_2$RM in the negative 
coupling regime is interpreted as the remnant of the $\lambda=0$ singularity of 
SRC. There remains the interesting question if by allowing 
for more than two link lengths the phase transition of such extended 
$Z_2$RM approaches that of SRC. Then the situation might resemble the more 
involved four-dimensional case where one has to deal with 10 edges per 
simplex and the nontrivial Einstein-Hilbert action 
$\sum_{t\supset i}\delta_tA_t$ with 50 triangles $t$ per vertex $i$ in
Eq.~(\ref{Iq}). Thus the action $I(q)$ takes on a large variety of values 
already for $Z_2$RM and therefore SRC can be approximated more 
accurately.\cite{4d}

\end{document}